\documentclass{aa}  
\usepackage{graphicx}
\usepackage{txfonts}
\begin{document}
   \title{Mass loss out of close binaries}
    \subtitle{Case A RLOF}

   \author{W. Van Rensbergen, J.P. De Greve, N. Mennekens, K. Jansen \and C. De Loore}

   \offprints{W. Van Rensbergen}

   \institute{Astrophysical Institute, Vrije Universiteit Brussel, Pleinlaan 2, 1050 Brussels, Belgium\\
   \email {wvanrens@vub.ac.be}
   }

    \date{Received September 9, 2009; accepted November 18, 2009}

% \abstract{}{}{}{}{} 
% 5 {} token are mandatory
 
  \abstract
   {Matter leaving the donor during mass transfer spins up the gainer and creates a hot spot in the impact area. If the kinetic energy of the enhanced rotation combined with the radiative energy of the hot spot exceeds the binding energy of the system, matter can escape from the binary.}
   {We calculate the amount of mass lost during eras of fast mass transfer. We simulate the distribution of mass-ratios and orbital periods for interacting binaries with a B-type primary at birth where mass transfer starts during hydrogen core burning of the donor.}
   {We used the initial distributions of primary mass, mass-ratio and orbital period established in a previous paper. The amount of time the binary shows Algol characteristics within different values of mass-ratio and orbital period has been fixed from conservative and liberal evolutionary calculations. We use these data to simulate the distribution of mass-ratios and orbital periods of Algols with the conservative as well as the liberal model.}
   {Rapid rotation and hot spots are frequently observed at the surface of the gainer in a semi-detached binary. The mass transfer rate for low-mass binaries is never sufficiently large to achieve mass loss from the system. Intermediate-mass binaries blow away a large fraction of the transferred mass during short eras of rapid mass transfer.}
   {We compare mass-ratios and orbital periods of Algols obtained by conservative evolution with those obtained by our liberal model. We calculate the amount of matter lost according to our model by binaries with an early B-type primary at birth. Since binaries with a late B-type primary evolve almost conservatively, the overall distribution of mass-ratios will only yield a few Algols more with high mass-ratios than conservative calculations do. Whereas the simulated distribution of orbital periods of Algols fits the observations well, the simulated distribution of mass-ratios produces always too few systems with large values.}

   
    \authorrunning{W. Van Rensbergen et al.}  
     \titlerunning{Mass loss out of close binaries}
   \maketitle
%________________________________________________________________

\section{Introduction}

Eggleton (\cite{Eggleton}) introduced the denomination $liberal$ to distinguish binary evolution with mass and subsequent angular momentum loss from the $conservative$ case, where no mass leaves the system. A catalogue containing nowadays 240 conservative evolutionary tracks can be found at the VUB-website (\cite{VUB}). Van Rensbergen et al. (\cite{Walter2}) developed a liberal scenario in which mass can be lost from a binary during a short era of rapid mass transfer soon after the onset of Roche Lobe Overflow (RLOF). The joint online catalogue with currently 356 liberal evolutionary tracks is available at the Centre de Donn\'ees Stellaires (CDS). The grid of calculations covers only binaries with a B-type primary at birth and initial orbital periods so that RLOF starts during hydrogen core burning of the donor: the case A of RLOF. Systems undergoing a second era of RLOF after the onset of hydrogen burning in the shell of the donor are designated as cases A/B.
      
%__________________________________________________________________

\section{Initial conditions for binaries with a B-type primary}
\label{sec_Initial}

%                                     Two column figure (place early!)
%______________________________________________ Gamma_1 (lg rho, lg e)
%

Van Rensbergen et al. (\cite{Walter1}) used non-evolved systems in the 9th Catalogue of Spectroscopic Binaries of Pourbaix et al. (\cite{Pourbaix}) to establish the initial conditions for the evolution of binaries with a B-type primary at birth. We distinguish between late B-type primaries in the mass-range [2.5-7] $M_{\odot}$ and early B-type primaries in the range [7-16.7] $M_{\odot}$. The subscript $d$ is used throughout the paper for the binary component which is the donor during RLOF, whereas the subscript $g$ is used for its mass-gaining companion. The following initial conditions were found:

\begin{flushleft}
$\bullet$~~future donors follow a normalized IMF: $\xi$($M_{d}$)=C$\times$$M_{d}^{-\alpha}$~~with $\alpha$=2.35 as given by Salpeter  (\cite{Salpeter}).\\ 
$\bullet$~~initial orbital periods obey a normalized distribution: $\Pi$(P)=$\frac{A}{P}$ as given by Popova et al. (\cite{Popova}). Different distributions between small and large initial periods (Van Rensbergen et al. \cite{Walter1}) do not have to be taken into account since we consider only small periods leading to RLOF A in this paper.\\
$\bullet$~~initial mass-ratios (q=${M_{g}\over M_{d}}$) follow a normalized distribution: $\Psi$(q)=C$\times$$(1+q)^{-\delta}$~;~$\delta$=0.65 for non-evolved binaries with a late B primary and $\delta$=1.65 for non-evolved binaries with an early B primary, as given by Van Rensbergen et al. (\cite{Walter1}).\\  
\end{flushleft}

\section{Amount of mass driven out of a close binary}
\label{sec_Amount}

Using a Monte-Carlo simulation starting from a large number of non-evolved binaries, the liberal binary evolutionary scenario of Van Rensbergen et al. (\cite{Walter2}) has been applied to binaries with a B-type primary at birth and undergoing RLOF A, i.e. during hydrogen core burning of the donor. The details used in the calculations can be found in that paper. 

\subsection{The contact phase}

A hot spot is created on the gainer's equator or on the edge of its accretion disk when the binary is semi-detached. For initial mass-ratios around 0.4 or lower (and in some cases for mass-ratios as high as 0.6), systems undergoing RLOF A evolve into contact soon after the onset of mass transfer, as a result of a rapid increase of the outer surface layers of the gainer. The effect of an energy stream during contact has been studied by Nakamura $\&$ Nakamura (\cite{Nakamura1},\cite{Nakamura2}) and Packet (\cite{Packet}, available upon request). Packet (\cite{Packet}) found that the effect has negligible influence on the evolution of the binary if the energy stream goes from gainer to loser (i.e. if the gainer is the hotter star, which is the case in our systems). Hence, we assume that mass transfer continues during this phase, without any extra energy transfer.

No hot spot can be created or spin-up generated during contact since the geometrical impact parameter of the system vanishes in that case. The binary suffers then no mass-loss and does not meet the Algol requirements. The contact phase does not last long and the system returns into the semi-detached state soon after mass-ratio reversal, as a combined result of a decrease of the mass transfer rate and the increase of the distance between the components. Initial mass-ratios smaller than 0.25 have been excluded from our statistical analysis because we assume that contact in those cases will provoke merging of the system. Initial mass-ratios larger than 0.25 undergoing contact have been included for the reasons given in this subsection and also because their exclusion would lead to the production of a too small number of Algol type binaries.

\subsection{The hot spot}

The accuracy of the semi-detached scenario is restricted by the value of the radiative efficiency of accretion $\tilde{K}$ which defines the quantity $L_{acc}$$\times$$\tilde{K}$ exerting the radiation pressure of a hot spot. $L_{acc}$ is that part of $L_{acc}^{\infty}$ = G$\times \frac {M_{g}\times M_{d}^{RLOF}}{R_{g}}$ that is available after reduction due to the fact that matter impinging on the gainer starts at the first Lagrangian point and not at infinity. This accretion luminosity is weakened by the fact that only a fraction can be converted into radiation and strengthened because the energy of the impacting material is concentrated in a hot spot which is significantly smaller than the entire gainer's surface. Van Rensbergen et al. (\cite{Walter2}) defined a quantity $K$ which enables the calculation of the contribution of the hot spot to the total luminosity of the gainer. It is easier to visualize the action of the hot spot using $\tilde{K}$=$\frac{1}{K}$ as given by relation (\ref{Ktilde}):

 \begin{equation}
 \tilde{K}~=~\frac{1}{K}~=~{{{\lbrack \frac {R_{g}} {R_{\odot}} \rbrack}^2}\times {\lbrack T_{spot}^{4}-T_{eff,g}^{4} \rbrack} \over \frac{L_{acc}}{L_{\odot}}\times (5770)^{4}}
 \label{Ktilde} 
\end{equation}

Unfortunately, there are only 11 reliable hot spot temperatures available in the literature. Eight systems (VW Cep, CN And, KZ Pav, V361 Lyr, RT Scl, U Cep, U Sge, and SV Cen) are direct impact systems, whereas three of them (SW Cyg, V356 Sgr and $\beta$ Lyr) have a transient accretion disk. In the case of the formation of a hot spot on the edge of an accretion disk we have to replace in equation (\ref{Ktilde}): $R_{g}$ by $R_{disk}$ and $T_{eff,g}$ by $T_{edge,disk}$. Our liberal evolutionary calculations have thus been gauged with small number statistics, producing an empirical relation for the radiative efficiency of accretion $\tilde{K}$ concentrated in a hot spot, increasing with the total mass of the system as: 

 \begin{equation}
 \tilde{K} = 4.386 \times {\lbrack{\frac{M_{d}}{M_{\odot}}+\frac{M_{g}}{M_{\odot}}\rbrack}}^{1.735}
 \label{Ktildenumber} 
\end{equation}

Radiative efficiency, $\eta$, is usually defined through the luminosity $L_{add}$ which is added to a system as a consequence of the transfer of matter at a rate $\dot{M}$:

\begin{equation}
 L_{add}=\eta~\dot{M}~c^{2}
 \label{eta} 
\end{equation}

A numerical value can hence be calculated from relation (\ref{Ktildenumber}) for the radiative efficiency of mass accretion by a Main Sequence gainer:

\begin{equation}
 \eta~= 1.88\times{10^{-5}}\times {\lbrack{\frac{M_{d}}{M_{\odot}}+\frac{M_{g}}{M_{\odot}}\rbrack}}^{1.735}\times{D}\times{S}
 \label{etanumber} 
\end{equation}

The factor $D$ is the geometric factor taking into account that matter does not fall onto the gainer from infinity but from the first Lagarangian point. This factor is zero for a contact system and goes to unity as $L_{1}$ goes to infinity. The factor $S$ is the fractional surface area of the hot spot. Consequently we can compare the values of $\eta$ calculated with relation (\ref{etanumber}) with values that are well known from fundamental physics: $\eta$ = 0.007 for complete nuclear fusion of hydrogen, $\eta$ grows smoothly from 0.057 for a mass gaining non-rotating black hole to 0.32 for a black hole rotating at maximum plausible spin (Thorne~\cite{Thorne}).

 \subsection{Tidal interaction}
  
All the systems in the grid have been calculated both with strong and weak tidal interaction. The formalism for the tidal interaction was taken from Zahn (\cite{Zahn}), who gives a suitable approximation for the synchronisation time-scale: 

 \begin{equation}
 \tau_{sync}~(yr)={q^{-2}} \times {\lbrack {a\over R_{g}} \rbrack} ^{6}
\label{tausync} 
  \end{equation}
  
 This expression uses the semi major axis $a$ of the binary and a mass-ratio $q$, in which the star that has to be synchronized is in the denominator. This is the $gainer$ in our case, so that ${q}={M_{d}\over M_{g}}$.

Tidal interactions modulate the angular velocity of the gainer $\omega_{g}$ with the angular velocity $\omega_{orb}$ of the system. According to Tassoul (\cite{Tassoul}) one can write:

 \begin{equation}
{1 \over {\omega_{g}-\omega_{orb}}} \times {d \omega_{orb} \over  d t}= - {1\over {\tau_{sync} \times f_{sync}}}= -{1 \over t_{sync}}
\label{fsyncsync} 
  \end{equation}
  
Tidal interactions spin the gainer down when $\omega_{g} > \omega_{orb}$. Tides spin the gainer up when $\omega_{g} < \omega_{orb}$. $f_{sync}$ = 1 represents weak tidal interactions whereas $f_{sync}$ = 0.1 implies strong tides.

\begin{figure*}[!ht]
\centering
\includegraphics[width=9.6cm]{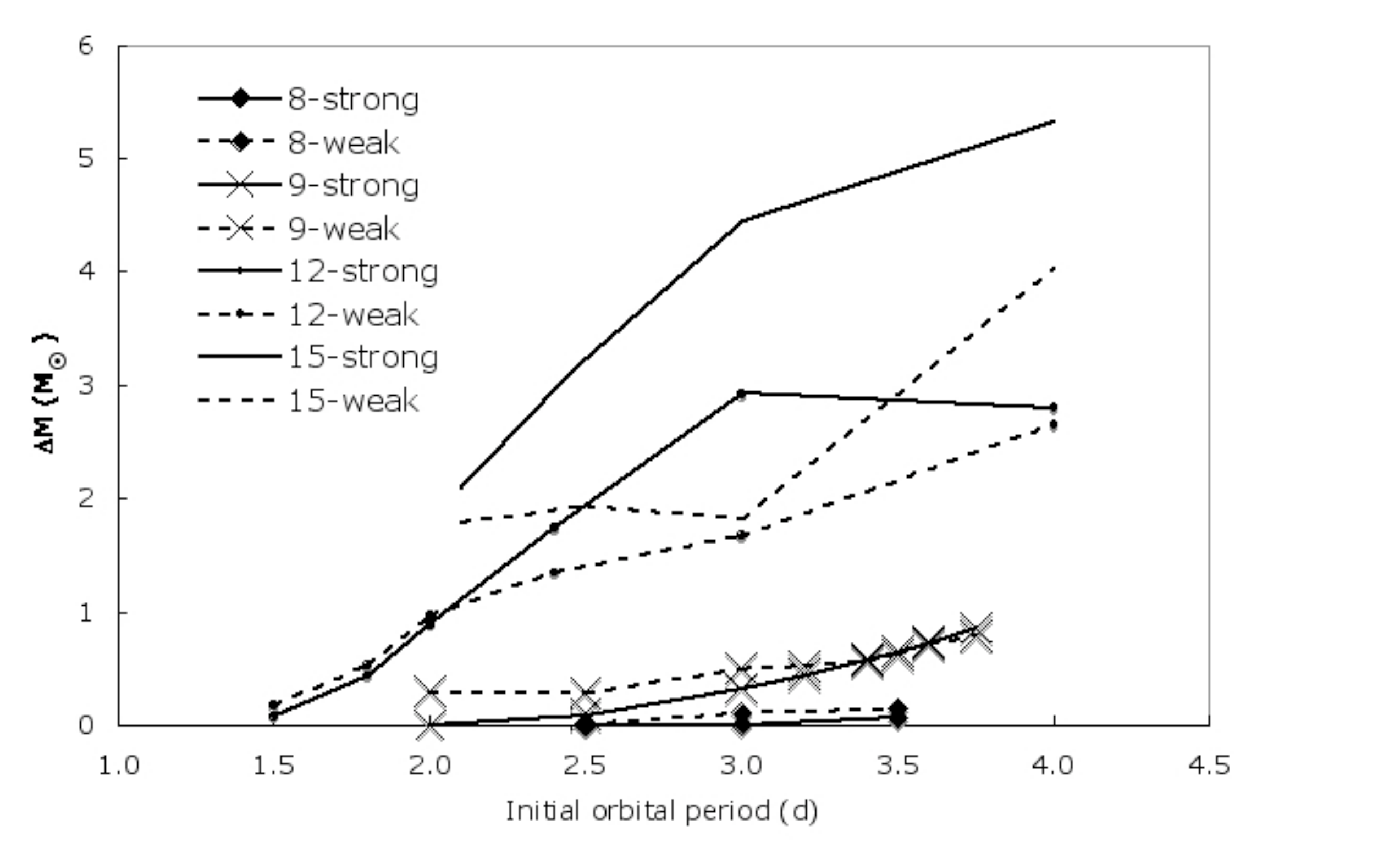}
\caption{Amount of mass lost by a binary with a B-type primary at birth and an initial mass-ratio ${M_{g}\over M_{d}}$ = 0.4. The amount of mass (in $M_{\odot}$) is on the vertical axis. The initial period is on the horizontal axis. The initial mass of the donor and the type of tidal interaction are mentioned in the legend. Donors with initial masses below 8 $M_{\odot}$ lose only a little amount of mass and are not included.} 
\label{fig_fig1}
\end{figure*}

\begin{figure*}[!ht]
\centering
\includegraphics[width=9.6cm]{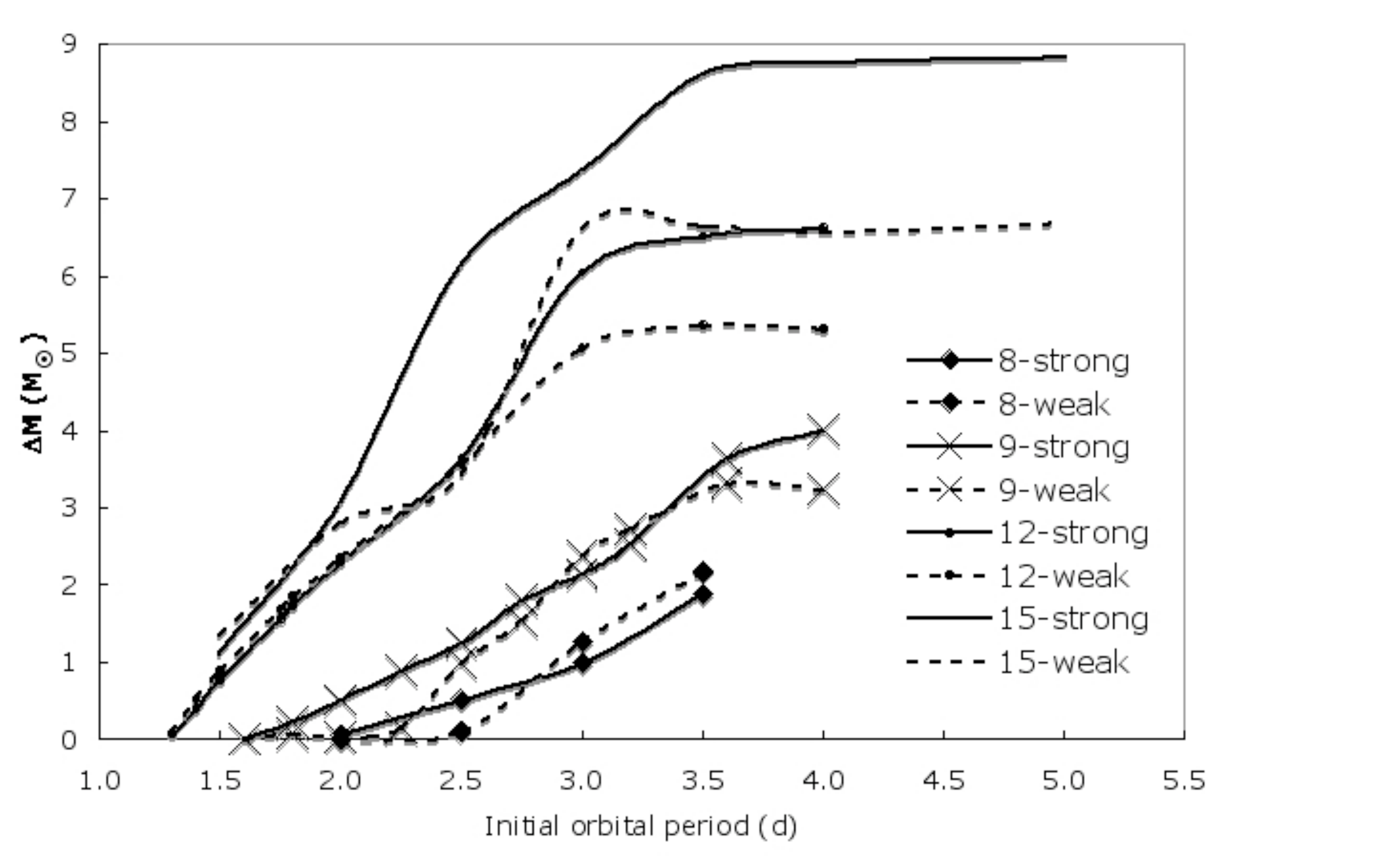}
\caption{Amount of mass lost by a binary with a B-type primary at birth for ${M_{g}\over M_{d}}$ = 0.6. The axes have the same meaning as in Figure $\ref{fig_fig1}$.}
\label{fig_fig2}
\end{figure*}

\begin{figure*}[!ht]
\centering
\includegraphics[width=9.6cm]{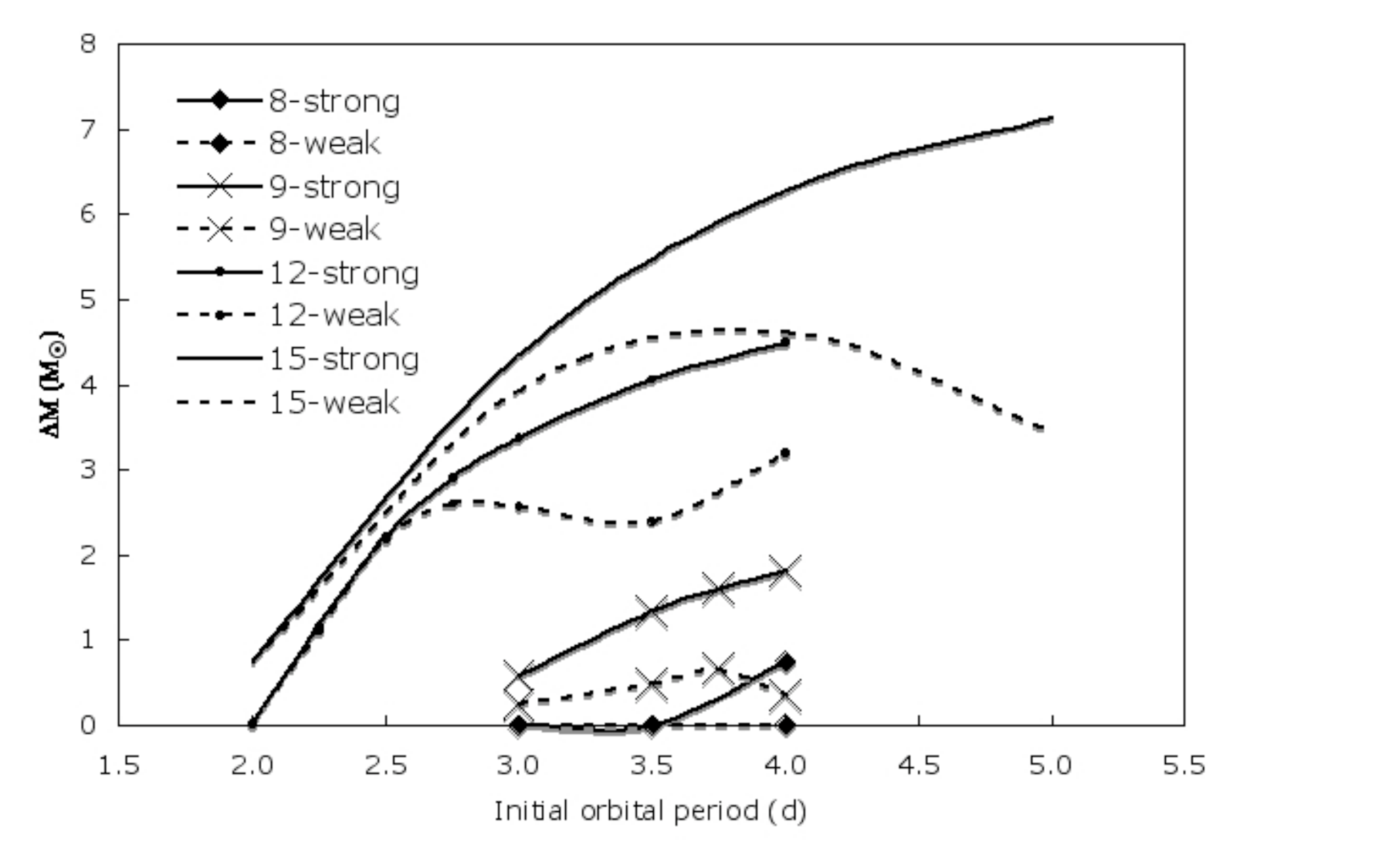}
\caption{Amount of mass lost by a binary with a B-type primary at birth for ${M_{g}\over M_{d}}$ = 0.9. The axes have the same meaning as in Figure $ \ref{fig_fig1}$.}
\label{fig_fig3}
\end{figure*}

Strong tidal coupling should be preferable when the spherical shape of the gainer is severely elongated due to rapid rotation. As a result we found that binaries may lose matter during short eras of rapid RLOF which occur soon after the onset of RLOF during hydrogen core burning of the donor and sometimes also short after the subsequent onset of RLOF during hydrogen shell burning of the donor. Binaries with initial primary masses below 6 $M_{\odot}$ certainly show spinning up of the gainer and increased accretion luminosities concentrated in hot spots, but the combined energies of both events are never sufficient to overcome the binding energy of the system: those systems evolve conservatively. Following increasing values of the initial primary mass we find that binaries with initial primary masses below 8$M_{\odot}$ do not lose a significant fraction of the transferred mass. Figures \ref{fig_fig1} to \ref{fig_fig3} show the amounts of mass lost by systems with initial primary masses from 8$M_{\odot}$ on, respectively for ${M_{g}\over M_{d}}$= 0.4, 0.6 and 0.9. Up to 9$M_{\odot}$ are lost by systems with a 15$M_{\odot}$ primary mass at birth. There are no huge differences in mass loss between similar cases calculated respectively with weak and strong tidal interaction.

\section{The distribution of mass-ratios and orbital periods of Algols}
\label{sec_section4}

\subsection{The observed distributions}
\label{sec_obs}

Van Rensbergen et al. (\cite{Walter2}) used the mass-ratio and orbital period distribution of Algols out of a sample of 303 observed systems. These systems have been taken from the catalogue of Budding et al. (\cite{Budding et al.}) extended with semi-detached Algols from the catalogue of Brancewicz et al. (\cite{Brancewicz et al.}). All these systems are issued from a binary with a B-type primary at birth. A majority of 268 systems have a late B-type primary progenitor, leaving only 35 systems with an early B-type primary at birth. The initial conditions described in section (\ref{sec_Initial}) use q = ${M_{g}\over M_{d}}$ as a definition of the mass-ratio, but since the gainer has become the most massive component of the Algol-system, we use q = ${M_{d}\over M_{g}}$ as the definition of mass-ratio of an Algol-system. The observed Algols combine a large fraction of systems where Algol characteristics are produced from hydrogen core burning of the donor on (case A and case A/B) and a smaller fraction where Algols are produced only after the ignition of hydrogen shell burning in the donor (case B).

\subsection{The calculated distributions}
\label{sec_calc}

Our conservative and liberal binary calculations consider a binary to be an Algol when the semi-detached system shows the typical characteristics as mentioned by Peters (\cite{Peters}):

\begin{flushleft}
$\bullet$~The less massive donor fills its Roche Lobe.\\ 
$\bullet$~The most massive gainer does not fill its Roche lobe and is still on the Main Sequence\\
$\bullet$~The donor is the cooler, the fainter and the larger star\\  
\end{flushleft}

The calculations in this paper do not include the Case B-calculations, so that the final comparison between observation and theory will be delayed until the Case B-calculations are also included in the CDS-catalogue. The calculations in this paper however compare the results as obtained with the assumption of strong tidal interaction with those obtained with weak tides. Although significant differences frequently occur between both assumptions, they yield a very similar overall distribution of mass-ratios and orbital periods of Algols.

\subsection{The mass-ratio distribution of Algols with a late B-type primary at birth}
\label{sec_late}

\begin{figure*}[!ht]
\centering
\includegraphics[width=9.6cm]{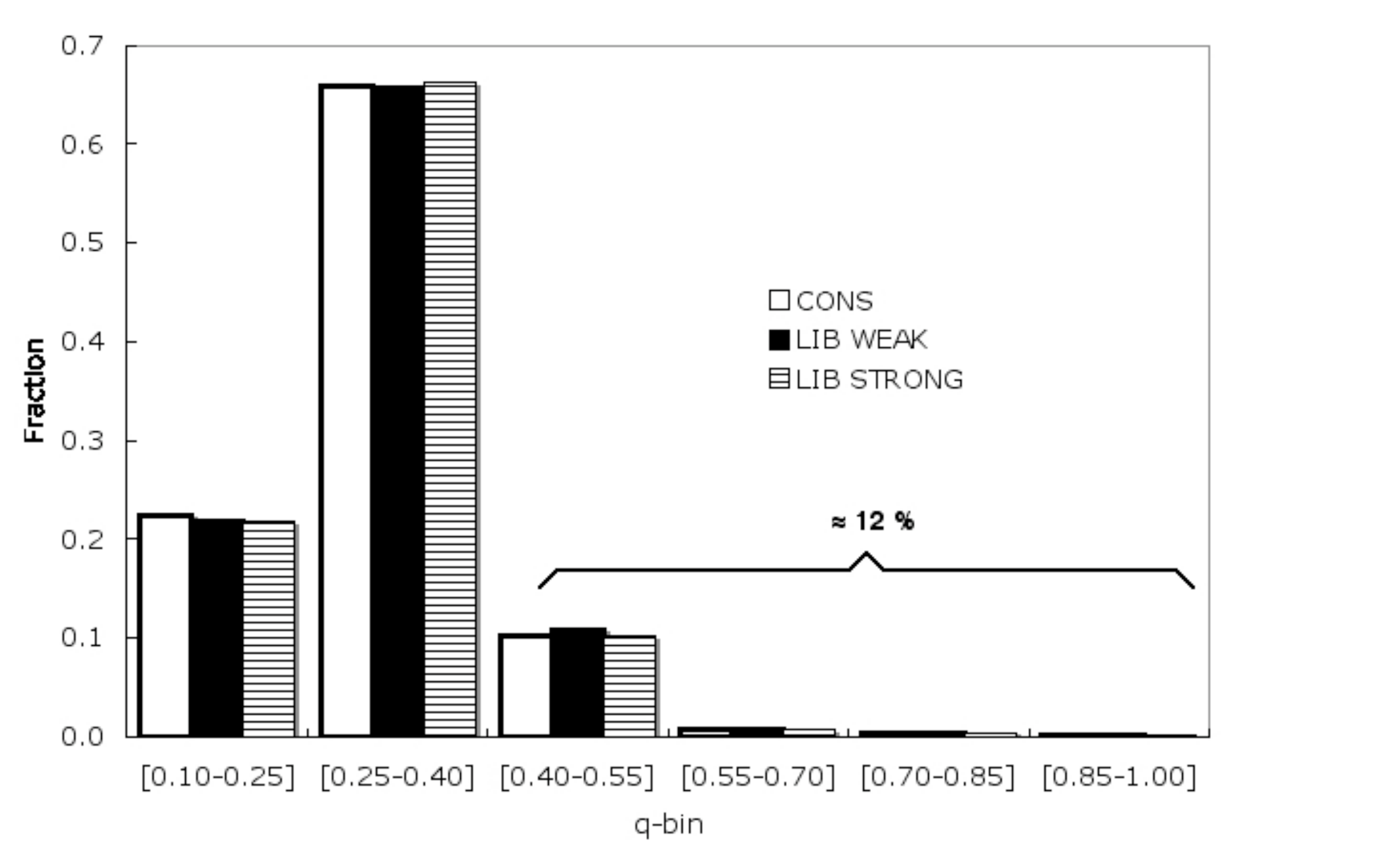}
\caption{Simulated distribution of mass-ratios of Algols issued from a binary with a late B-type primary at birth and an initial orbital period so that Case A RLOF occurs. Since these binaries do not lose a large amount of mass, the mass-ratio distribution is approximately the same for conservative evolution as compared to liberal evolution, producing only $\approx$ 12 $\%$ of Algols with high mass-ratios: $q$ $\in$ [0.4-1].} 
\label{fig_fig4}
\end{figure*}

Since binaries with a late B-type primary at birth hardly lose any mass during their evolution, the mass-ratio distribution as obtained with the liberal assumptions will not differ very much from the one obtained using the conservative binary evolutionary code. The fact that $\approx$ 45 $\%$ out of a sample of 268 Algols are observed with high mass-ratios ($q$ $\in$ [0.4-1]) is thus neither reproduced with the liberal binary evolutionary scenario nor with the conservative code. All these codes produce only $\approx$ 12 $\%$ of Algols with a mass-ratio q above 0.4. Figure {\ref{fig_fig4}} shows the mass-ratio distribution of Algols with a late B-type primary at birth as obtained from theory with the conservative and the two different liberal assumptions about the strength of the tidal interaction.

\subsection{The mass-ratio distribution of Algols with an early B-type primary at birth}
\label{sec_early}

\begin{figure*}[!ht]
\centering
\includegraphics[width=9.6cm]{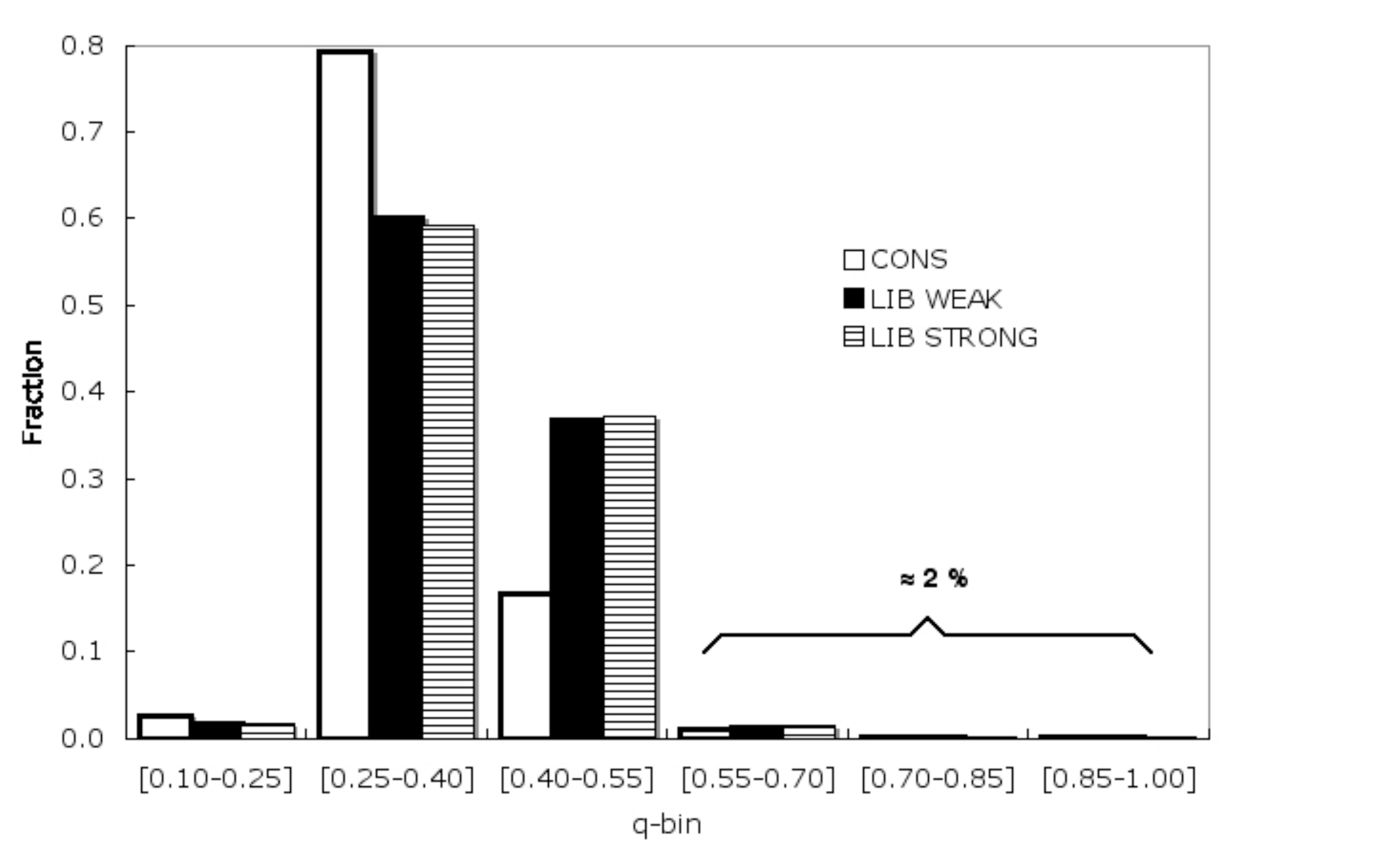}
\caption{Simulated distribution of mass-ratios of Algols issued from a binary with an early B-type primary at birth and initial orbital periods so that Case A RLOF occurs. These binaries lose a large amount of mass so that the mass-ratio distribution differs for conservative evolution as compared to liberal evolution, producing $\approx$ 39 $\%$ of Algols with high mass-ratios ($q$ $\in$ [0.4-1]) in the case of liberal evolution only.} 
\label{fig_fig5}
\end{figure*}

In section (\ref{sec_Amount}) we showed that binaries with an early B-type primary at birth lose a large fraction of the transferred mass during their evolution. The short liberal era during which mass is lost occurs when the mass transfer rate is large, i.e. soon after the onset of RLOF when the binary is in its pre- or early-Algol stage. The mass-ratio distribution as obtained with the liberal assumptions will now differ very much from the one obtained using the conservative binary evolutionary code. The fact that $\approx$ 46 $\%$ out of a sample of (only) 35 Algols are observed with high mass-ratios ($q$ $\in$ [0.4-1]) is reproduced with the liberal binary evolutionary scenario only ($\approx$ 39 $\%$). The large fraction of Algols observed with very high mass-ratios ($\approx$ 17 $\%$ with $q$ $\in$ [0.85-1]) remains however very hard to explain. Since the liberal era occurs mainly in the very early Algol-stage (when $q$ $\approx$ 1) it is clear that the Algol will populate the lower values of $q$ during the much longer lasting eras of quiet and slow mass transfer. The binary loses mass during the short lasting era of rapid mass transfer and then shows Algol characteristics during long lasting eras of quiet and slow mass transfer. Figure {\ref{fig_fig5}} shows the mass-ratio distribution of Algols with an early B-type primary at birth as obtained from theory with the conservative and the two different liberal assumptions. These conclusions are weakened by the fact that they are based on a comparison with only 35 Algols.

\subsection{Period distribution of Algols with a B-type primary at birth}
\label{sec_period}

\begin{figure*}[!ht]
\centering
\includegraphics[width=9.6cm]{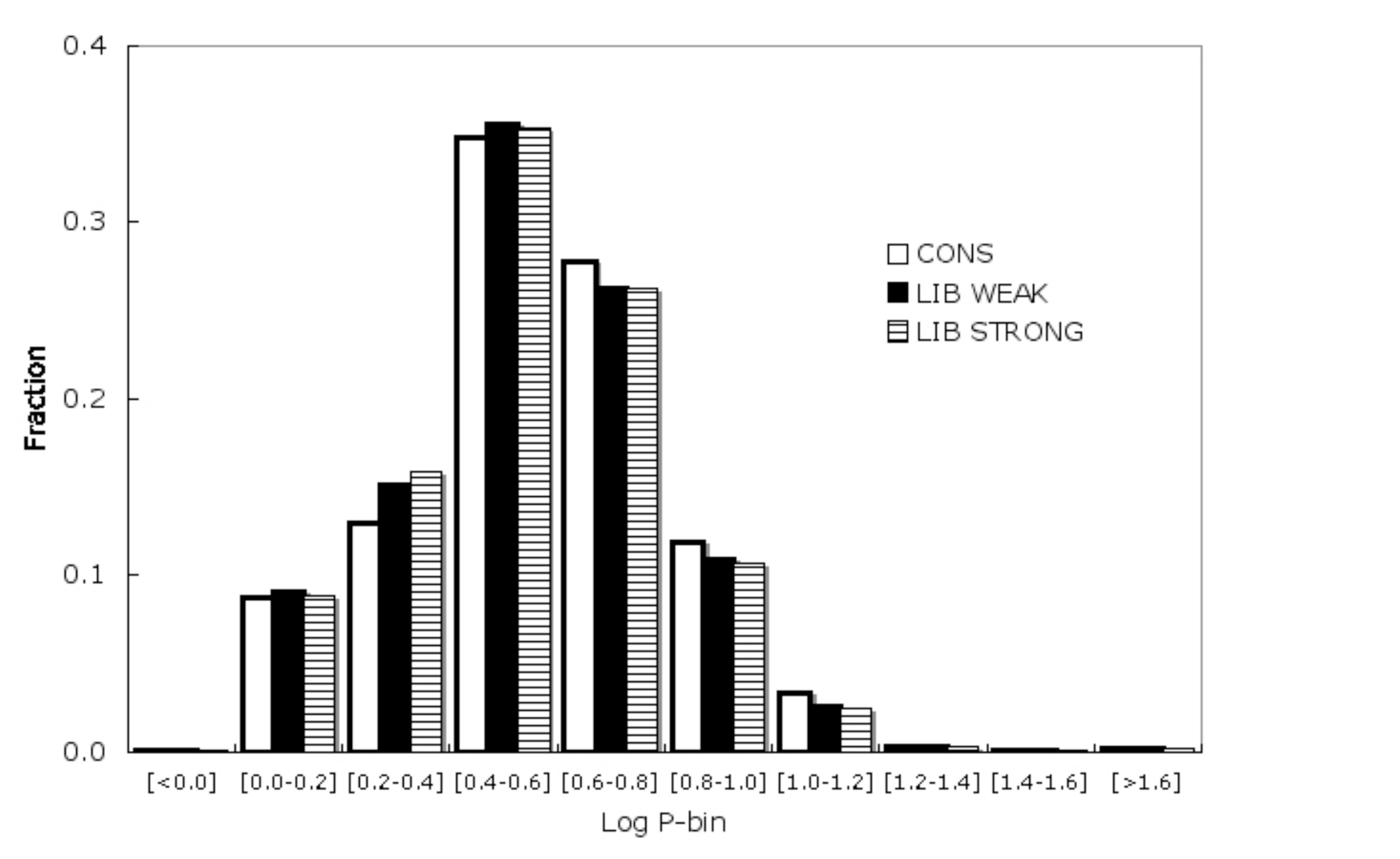}
\caption{Simulated distribution of orbital periods of Algols issued from a binary with a B-type primary at birth and initial orbital periods so that Case A RLOF occurs. Conservative and liberal evolution produce almost the same result.}
\label{fig_fig6}
\end{figure*}

Figure {\ref{fig_fig6}} shows the orbital period distribution of Algols with an early B-type primary at birth as obtained from theory with the conservative and the two different liberal assumptions. We find no large differences between the conservative and liberal results. It will be very hard to reproduce by theory the Algols observed with orbital periods below 1 day ($\approx$ 9 $\%$) since those systems always merge. Algols observed with orbital periods above 15 days ($\approx$ 5 $\%$) will be reproduced by theory when one includes those Algols which are only formed after ignition of hydrogen in the shell of the donor. It was e.g. illustrated for the conservative case by Van Rensbergen (\cite{Walter3}) that these cases B produce a large fraction of Algols with large orbital periods. These Algols will however not influence the final global distribution of mass-ratios and orbital periods of Algols very much, since their contribution to the Algol-population will not be very large due to the fact that the hydrogen shell burning duration will always last much shorter than its previously lived hydrogen core burning time. Anticipating on finishing our catalogue of binary evolutionary calculations with the cases B of RLOF, we may conclude that theory reproduces the observations well.

\section{Conclusions}

Mass impinging from the donor spins the gainer up and creates a hot spot in its equatorial zone or on the edge of its accretion disk. The combined energy of the enhanced rotation and increased radiation from the hot spot may exceed the binding energy of the system. We present the results of liberal evolutionary calculations for binaries with a B-type primary at birth and with small initial orbital periods so that RLOF starts during hydrogen core burning of the donor. A significant fraction of the transferred mass is lost by the system in the case of binaries with an early B-type primary at birth. Systems with a late B-type primary at birth hardly lose any matter. 

The observed distribution of orbital periods of Algols is well reproduced by conservative as well by liberal theoretical calculations.

Although the theoretically calculated liberal mass-ratio distribution of Algols with an initial early B-type primary fits the observations much better, the overall observed mass-ratio distribution of Algols still shows too many systems with large mass-ratios. This is due to the fact that the calculated era of rapid mass transfer is very short so that the binary rushes through the states with large mass-ratios. Binary evolutionary calculations yielding eras of rapid mass transfer lasting for a longer time with somewhat lower peak values of the mass transfer rate would produce more Algols with large mass-ratios but have never been published. We have compared the mass transfer rates as obtained by our binary evolutionary code in the conservative mode with mass transfer rates for conservative evolution as produced by previous authors. According to Kippenhahn et al. (\cite{Kippenhahn1}, \cite{Kippenhahn2}) a 9 $M_{\odot}$ donor transfers more than 5 $M_{\odot}$ to his initial 5 $M_{\odot}$ gainer in 6 $\times$ $10^4$ years during hydrogen core burning of the donor and almost 7 $M_{\odot}$ in 4 $\times$ $10^4$ years when RLOF starts after exhaustion of hydrogen in the core of the donor. A 2 $M_{\odot}$ donor transfers 0.45 $M_{\odot}$ to his 1 $M_{\odot}$ companion in 3.1 $\times$ $10^5$ years in the rapid phase of mass transfer during hydrogen core burning of the donor. Paczy\'nski et al. (\cite{Paczynski1}, \cite{Paczynski2}) calculated the conservative evolution of a binary with a 16 $M_{\odot}$ future donor at birth and a 10.67 $M_{\odot}$ companion. With an initial orbital period leading to case A RLOF he finds that almost 8 $M_{\odot}$ are transferred to the gainer in 4 $\times$ $10^4$ years. When RLOF starts after exhaustion of hydrogen in the core of the donor a short era of mass of mass transfer is found with a peak value as high as 3.4 $\times$ $10^{-4}$ $M_{\odot}\over year$. Our calculated durations of rapid mass transfer are very similar to those mentioned above whereas our peak values are somewhat lower. This is due to the fact that our stellar models are calculated with Rogers- Iglesias opacities (\cite{Rogers}) which were not available previously. Therefore, our calculated durations and peak values of mass transfer rates agree very well with those as published by Nelson and Eggleton (\cite{Nelson}) for a representative set of interacting binaries. The occurrence of many observed Algols with large mass-ratios thus remains unexplained. Future investigations should explore other interactions between the gravitational RLOF and the internal thermal structure driving the evolution of the radius of the donor.

Podsiadlowki et al. (\cite{Podsiadlowski}) pointed out that liberal theoretical calculations depend very much on the amount of mass lost from the system (characterized by the parameter $\beta$) and the amount of angular momentum taken away by this matter (characterized by the parameter $\alpha$). Our liberal code calculates $\beta$(t) self-consistently within the model and assumes that matter is lost from the hot spot on the gainer (or edge of its accretion disk) so that the escaping matter takes only the angular momentum of the gainer's orbit. It is clear that if matter would escape at another location (another choice of the parameter $\alpha$, e.g. characteristic for $L_{2}$ as the position of mass loss from the system) the calculated population of Algols could be different.

\begin{acknowledgements}
      We thank Peter Eggleton and the anonymous referee for their comments and suggestions.
      \end{acknowledgements}

\end{document}